\documentclass[twocolumn]{aastex62}
\graphicspath{{./}{figures/}}

\received{...}
\revised{...}
\accepted{\today}
\submitjournal{ApJ Letters}

\shorttitle{Upper limit on optical bursts from SGR J1935+2154}
\shortauthors{Zampieri et al.}

\begin{document}

\title{Deep upper limit on the optical emission during a hard X-ray burst from the magnetar SGR J1935+2154}

\correspondingauthor{Luca Zampieri}
\email{luca.zampieri@inaf.it}

\author[0000-0002-6516-1329]{Luca Zampieri}
\affil{INAF - Osservatorio Astronomico di Padova, Vicolo dell'Osservatorio 5, 35122, Padova, Italy}

\author{Sandro Mereghetti}
\affiliation{INAF - Istituto di Astrofisica Spaziale e Fisica Cosmica, Via Alfonso Corti 12, 20133, Milano, Italy}

\author{Roberto Turolla}
\affiliation{Dipartimento di Fisica ed Astronomia, Universit\`a di Padova, Via Francesco Marzolo 8, 35131, Padova, Italy}
\affiliation{Mullard Space Science Laboratory, University College London, Holmbury St Mary, Dorking, Surrey RH5 6NT, UK}

\author{Giampiero Naletto}
\affiliation{Dipartimento di Fisica ed Astronomia, Universit\`a di Padova, Via Francesco Marzolo 8, 35131, Padova, Italy}
\affiliation{INAF - Osservatorio Astronomico di Padova, Vicolo dell'Osservatorio 5, 35122, Padova, Italy}

\author{Paolo Ochner}
\affiliation{Dipartimento di Fisica ed Astronomia, Universit\`a di Padova, Via Francesco Marzolo 8, 35131, Padova, Italy}
\affiliation{INAF - Osservatorio Astronomico di Padova, Vicolo dell'Osservatorio 5, 35122, Padova, Italy}

\author{Aleksandr Burtovoi}
\affiliation{INAF - Osservatorio Astrofisico di Arcetri, Largo Enrico Fermi 5, 50125 Firenze, Italy}
\affiliation{INAF - Osservatorio Astronomico di Padova, Vicolo dell'Osservatorio 5, 35122, Padova, Italy}

\author{Michele Fiori}
\affiliation{Dipartimento di Fisica ed Astronomia, Universit\`a di Padova, Via Francesco Marzolo 8, 35131, Padova, Italy}
\affiliation{INAF - Osservatorio Astronomico di Padova, Vicolo dell'Osservatorio 5, 35122, Padova, Italy}

\author{Cristiano Guidorzi}
\affiliation{Dipartimento di Fisica e Scienze della Terra, Universit\`a di Ferrara, Via Saragat 1, 44122 Ferrara, Italy}
\affiliation{INFN - Sezione di Ferrara, Via Saragat 1, 44122 Ferrara, Italy}
\affiliation{INAF - Osservatorio di Astrofisica e Scienza dello Spazio di Bologna, Via Piero Gobetti 101, 40129 Bologna, Italy}

\author{Luciano Nicastro}
\affiliation{INAF - Osservatorio di Astrofisica e Scienza dello Spazio di Bologna, Via Piero Gobetti 101, 40129 Bologna, Italy}

\author{Eliana Palazzi}
\affiliation{INAF - Osservatorio di Astrofisica e Scienza dello Spazio di Bologna, Via Piero Gobetti 101, 40129 Bologna, Italy}

\author{Maura Pilia}
\affiliation{INAF - Osservatorio Astronomico di Cagliari, Via della Scienza 5, 09047 Selargius, Italy}

\author{Andrea Possenti}
\affiliation{INAF - Osservatorio Astronomico di Cagliari, Via della Scienza 5, 09047 Selargius, Italy}



\begin{abstract}

In September 2021 the magnetar SGR J1935+2154 entered a stage of burst/flaring activity in the hard X-ray band. On September 10, 2021 we observed SGR J1935+2154 with the fiber-fed fast optical photon counter IFI+Iqueye, mounted at the 1.22 m Galileo telescope in Asiago. During one of the IFI+Iqueye observing windows a hard X-ray burst was detected with the Fermi Gamma-ray Burst Monitor. We performed a search for any significant increase in the count rate on the 1-s, 10-ms and 1-ms binned IFI+Iqueye light curves around the time of the Fermi burst. No significant peak was detected with a significance above 3$\sigma$ in an interval of $\pm$90 s around the burst. Correcting for interstellar extinction ($A_V \simeq 5.8$ mag), the IFI+Iqueye upper limits to any possible optical burst from SGR J1935+2154 are $V=10.1$ mag, $V=7.2$ mag and $V=5.8$ mag for the 1-s, 10-ms and 1-ms binned light curves, respectively. The corresponding extinction corrected upper limits to the fluence (specific fluence) are $3.1 \times 10^{-10}$ erg cm$^{-2}$ (0.35 Jy s), $4.2 \times 10^{-11}$ erg cm$^{-2}$ (4.8 Jy $\cdot$ 10 ms), and $1.6 \times 10^{-11}$ erg cm$^{-2}$ (17.9 Jy ms), orders of magnitude deeper than any previous simultaneous optical limit on a magnetar burst. The IFI+Iqueye measurement can also place a more stringent constraint to the spectral index of the optical to hard X-ray fluence of SGR J1935+2154, implying a spectrum steeper than $\nu^{0.64}$.
Fast optical timing observations of bursts associated with radio emission have then the potential to yield a detection.

\end{abstract}



\section{Introduction}
\label{sect:intro}

SGR J1935+2154 is a Galactic magnetar, a non-accreting neutron star powered mainly by magnetic energy dissipation (see \citealt{2015SSRv..191..315M,2015RPPh...78k6901T,2017ARA&A..55..261K} for recent reviews). This class of high energy sources is characterized by strong variability and recurrent bursts of hard X-rays/soft gamma-rays (hence the name soft gamma-ray repeaters, SGRs), that have typical duration $\lesssim$ 1 s and peak luminosity $\lesssim 10^{39-41}$ erg s$^{-1}$. Hyper-energetic bursts with peak luminosity reaching $10^{47}$ erg s$^{-1}$, known as giant flares, have also been emitted sporadically by a handful of magnetars (e.g. \citealt{2005Natur.434.1107P}).

SGR J1935+2154 was discovered through the detection of a short burst with the Neil Gehrels Swift Observatory/Burst Alert Telescope in 2014 \citep{2014GCN.16520....1S}. Follow-up observations in the X-ray band revealed that the source is a magnetar with spin period $P=3.25$ s and period derivative ${\dot P} = 1.43 \times 10^{-11}$ s s$^{-1}$ \citep{2016MNRAS.457.3448I}, leading to a characteristic age of 3.6 kyr and a dipole magnetic field of $2.2 \times 10^{14}$ G. A very faint (H $\sim 24$) near infrared counterpart was identified with the Hubble Space Telescope \citep{2018ApJ...854..161L}. No pulsed radio emission has been observed so far 
(e.g. \citealt{2021arXiv210604821T}),
although a tentative detection was recently reported \citep{2020ATel14084....1Z}.

The source is projected in the direction of the Galactic plane and could be associated with the supernova remnant G57.2+0.8 for which several distance estimates are reported in the literature 
(see e.g. \citealt{2020ApJ...905...99Z} and references therein).
In the following, we will adopt the distance derived from an analysis of the dust-scattering ring observed in X-ray images taken with the Swift X-ray Telescope instrument in 2020 ($d = 4.4_{-1.3}^{+2.8}$ kpc; \citealt{2020ApJ...898L..29M}), which is independent of the supernova remnant association.

SGR J1935+2154 has sporadically gone through phases of hard X-ray burst/flaring activity 
(e.g. \citealt{2020ApJ...902L...2B,2020ApJ...893..156L}).
The outburst episode observed in 2020 culminated in tens of bursts emitted in a few days
(see e.g. \citealt{2020ApJ...898L..29M} and references therein).
During this outburst, on 2020 April 28 an extremely bright millisecond-duration radio burst (FRB 200428) was emitted by SGR J1935+2154 and detected with the Canadian Hydrogen Intensity Mapping Experiment (CHIME; \citealt{2020Natur.587...54C}) and STARE2 \citep{2020Natur.587...59B} telescopes. The radio burst turned out to be temporally coincident with a bright hard X-ray burst detected with the INTEGRAL \citep{2020ApJ...898L..29M}, Konus-Wind \citep{2021NatAs...5..372R}, Insight Hard X-ray Modulation Telescope (HXMT; \citealt{2020Natur.587...63L}) and AGILE \citep{2021NatAs...5..401T} satellites. No optical counterpart was detected in a simultaneous observation performed with the BOOTES telescope \citep{2020Natur.587...63L} down to an extinction-corrected fluence of $\lesssim$4400 Jy ms. A near infrared campaign was also carried out with Palomar Gattini-IR observing system in the J band, placing an upper limit on the second-timescale extinction-corrected fluence of $\lesssim 125$ Jy ms \citep{2020ApJ...901L...7D}.

In the period between Sep 9 and Sep 22, 2021 SGR J1935+2154 entered a new stage of burst/flaring activity, detected with several satellites (GECAM, \citealt{2021GCN.30793....1X}; Konus-Wind, \citealt{2021GCN.30804....1R}; Fermi, \citealt{2021GCN.30806....1R}; AGILE, \citealt{2021GCN.30835....1U}). We then decided to target the source with our fast photon counter Iqueye \citep{2009A&A...508..531N}, which is fiber-fed at the 1.22 m Galileo telescope at the Asiago Astrophysical Observatory through the Iqueye Fiber Interface\footnote{https://web.oapd.inaf.it/zampieri/aqueye-iqueye/} (IFI; \citealt{2019CoSka..49...85Z}). Our main goal was exploiting the exquisite time resolution (up to ns) and high sensitivity of our instrumentation to search for second or sub-second optical flashes possibly associated to X-ray and/or radio bursts from SGR J1935+2154.



The observations presented here are part of a multiwavelength programme devoted to searching for prompt/delayed optical flashes from Fast Radio Bursts (FRBs) and Magnetars (e.g. \citealt{2020ApJ...896L..40P}). In Sect.~\ref{sect:observations} we present the observations of SGR J1935+2154 carried out on Sep 10, 2021 with IFI+Iqueye and the Fermi Gamma-ray Burst Monitor (GBM) instrument. In Sect.~\ref{sect:results} we show the results of the analysis, while in Sect.~\ref{sect:discussion} we discuss them within the framework of previous measurements.

\begin{table}
\caption{Log of the 2021 September unfiltered observations of SGR J1935+2154 carried out with IFI+Iqueye mounted at the 1.22 m Galileo telescope at the Asiago Astrophysical Observatory.}
\label{tab:log}
\centering
\begin{tabular}{l c c c}
\hline
Observation ID           &   Start time             &  Exposure        &   Rate    \\
                         &   (UTC)                  &  (s)             &   (kc/s)  \\
\hline
20210911-002926          &   2021-09-10 22:39:29.0  &  1197.4          &   1.8   \\
20210911-011727          &   2021-09-10 23:17:30.0  &  1797.6          &   2.0   \\
20210911-014851          &   2021-09-10 23:48:54.0  &  1797.6          &   2.3   \\
\hline
\end{tabular}
\end{table}

\section{Observations and data analysis}
\label{sect:observations}

\subsection{IFI+Iqueye}
\label{sect:ifi+iqueye}

On the night of September 10, 2021 we observed the area of the sky centered at the position of SGR J1935+2154 (RA=19 34 55.606, Dec=+21 53 47.45, J2000, error 0.2''; \citealt{2018ApJ...854..161L}) with the fiber-fed fast optical photon counter IFI+Iqueye \citep{2009A&A...508..531N,2019CoSka..49...85Z}, mounted at the 1.22 m Galileo telescope at the Asiago Astrophysical Observatory. The log of the observations, performed in white light without filters, and the average count rate (not background subtracted) measured at the position of SGR J1935+2154 are reported in Table~\ref{tab:log}. At the beginning and at the end of each acquisition a nearby sky region and a reference star close to the position of the source were also observed to monitor the quality and transparency of the sky. The first 10 minutes of the first observation (Obs. ID 20210911-002926) were affected by passage of slight veils and were then discarded, leaving a useful observing window of $\sim$20 minutes.

The target was carefully centered on the IFI instrument camera in such a way to match the position of the optical fiber injecting the light into Iqueye \citep{2019CoSka..49...85Z}. To this aim, an image of the field was previously acquired and astrometrically calibrated using nine stars in the field. The error of the target position registered on the image is 0.6'', significantly smaller than the optical fiber diameter. 

The photon event lists acquired with Iqueye were reduced using a dedicated software\footnote{QUEST v. 1.1.5, see \cite{2015SPIE.9504E..0CZ}.}.
Light curves binned at 1 ms, 10 ms and 1 s
were computed from the reduced event lists and searched for any significant rate increase.

\subsection{Fermi GBM}
\label{sect:fermi}

\begin{figure}
	\includegraphics[angle=0, width=0.94\columnwidth]{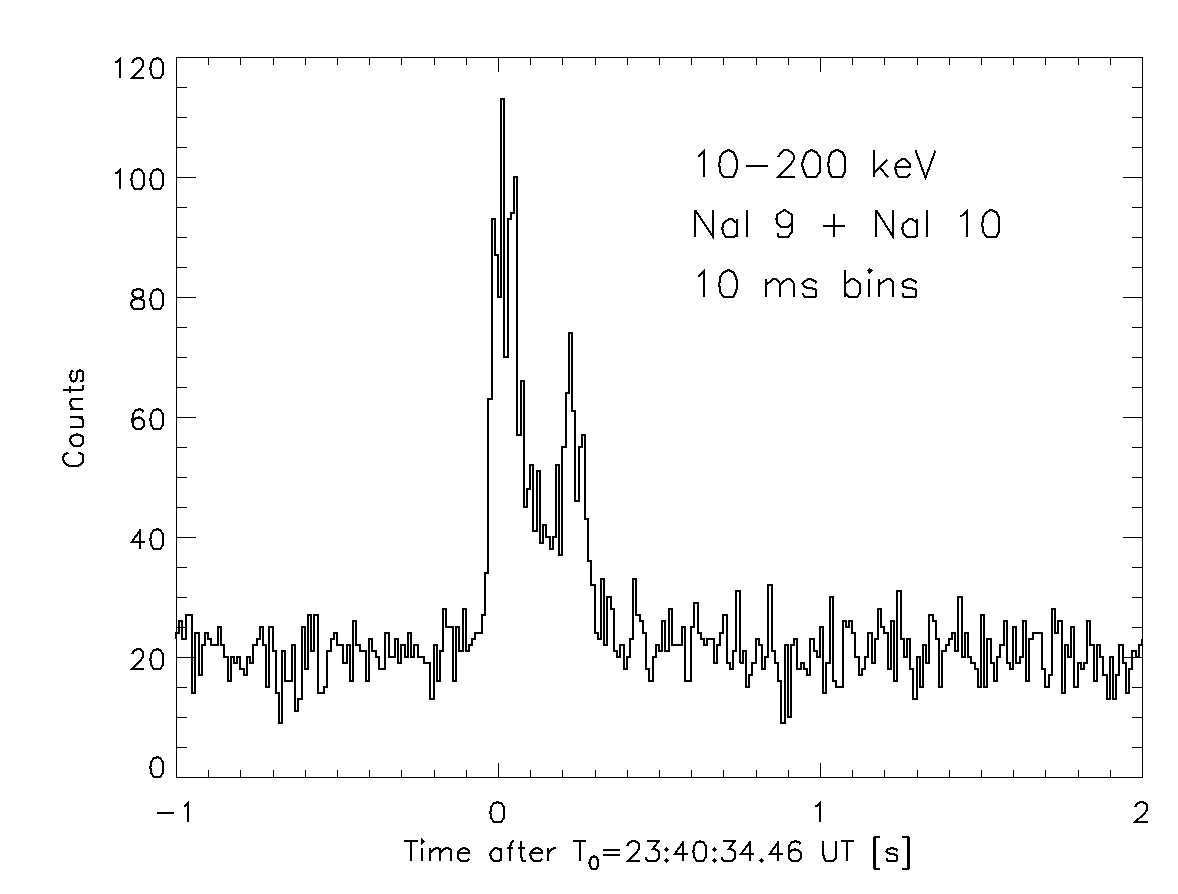}
    \caption{
    Fermi GBM light curve of the burst detected during the IFI+Iqueye observation. Time is measured from the burst trigger time T$_0$. The bin size is 10 ms.
    }
    \label{fig:gbm-lc}
\end{figure}

In September 2021 the Fermi GBM instrument revealed several bursts from the direction of SGR J1935+2154 \citep{2021GCN.30806....1R}. One of them occurred on September 10, during our IFI+Iqueye observations (GBM trigger n. 653010039, at T$_0$= 23:40:34.46 UTC). Figure~\ref{fig:gbm-lc} shows the 10-200 keV light curve of this burst binned at 10 ms, as derived by summing the counts of  the two NaI detectors with the best orientation for the magnetar direction (n. 9 and n. 10).

In order to measure the burst fluence, we performed a spectral analysis by extracting the spectra of the two GBM NaI detectors with the RMFIT software in the time interval  [T$_0 -0.04$ s, T$_0 + 0.38$ s], as well as those of the corresponding background. The latter was estimated by fitting with a linear function the count rates measured in time intervals of about 30 s before and after the burst.  We fitted the two spectra simultaneously in the energy range from 10 to 200 keV with an exponentially cut-off power law function defined as $N(E) = K E^{-\Gamma} e^{-(2-\Gamma)E/E_p}$ photons cm$^{-2}$ s$^{-1}$ keV$^{-1}$.

\noindent
A good fit was obtained with photon index $\Gamma=0.88^{+0.34}_{-0.38}$,   peak energy $E_p=23.5^{+2.2}_{-2.7}$ keV and flux $F=(7.2\pm0.4)\times10^{-7}$ erg cm$^{-2}$ s$^{-1}$ in the 10-200 keV range.  For a burst duration of  0.42 s, this corresponds to a fluence of $(3.0\pm0.2)\times10^{-7}$ erg cm$^{-2}$. 

\section{Results}
\label{sect:results}

\begin{figure}
        \includegraphics[angle=0, width=0.94\columnwidth]{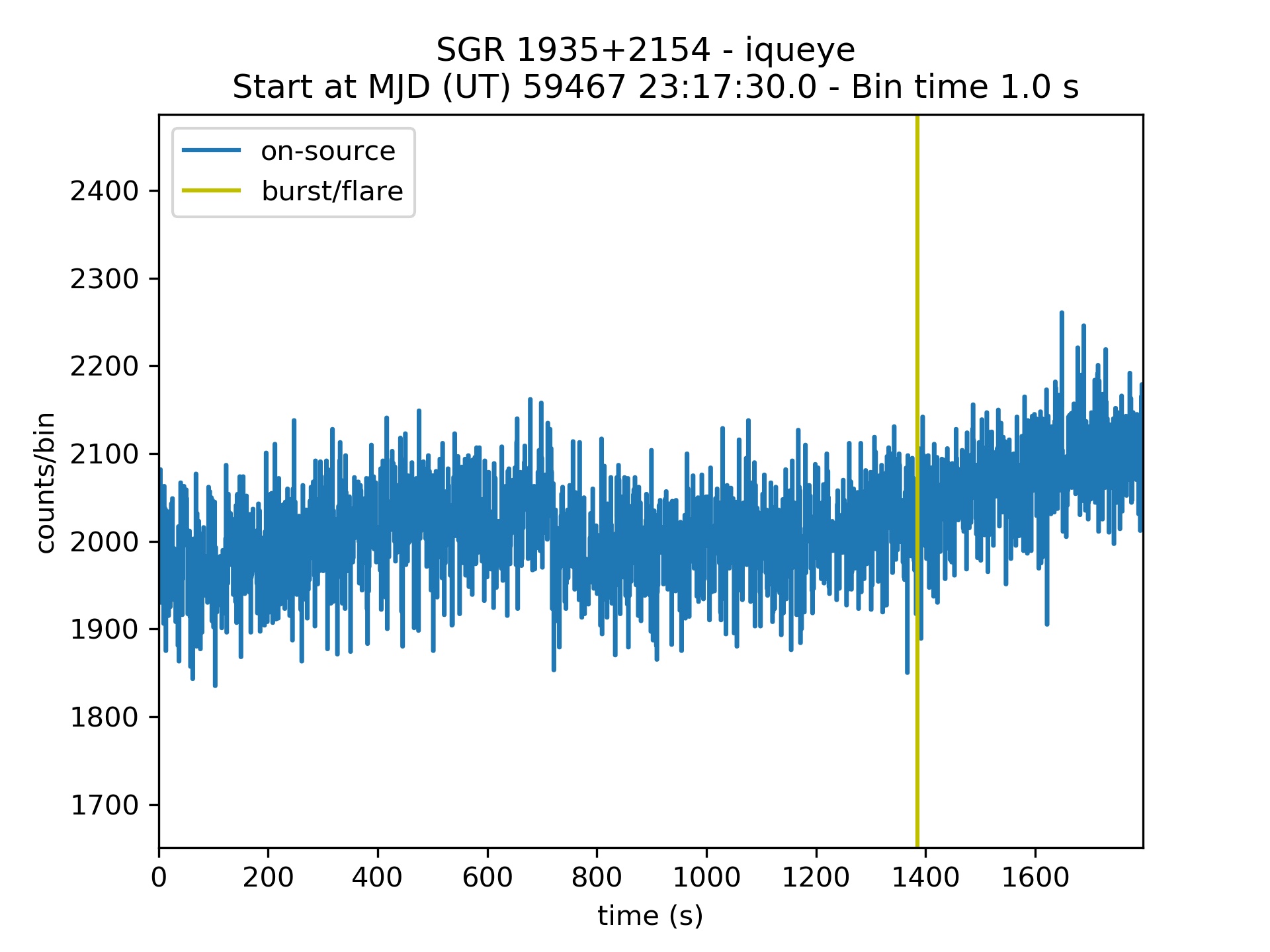}
    \caption{
    The 1-s binned light curve of the second IFI+Iqueye observation of SGR J1935+2154 performed on Sep 10, 2021 (Table~\ref{tab:log}, Obs. ID 20210911-011727). The minutes-timescale variations of the count rate are caused by changes of the background sky brightness. The yellow line marks T$_0$, the time of arrival of the Fermi GBM burst.
    }
    \label{fig:lciq}
\end{figure}

Figure~\ref{fig:lciq} shows the 1-s binned light curve of the second IFI+Iqueye observation.
We performed a search for any significant increase in the count rate on the 1-s, 10-ms and 1-ms binned optical light curves around the time of the Fermi GBM burst.

\begin{figure}
        \includegraphics[angle=0, width=0.94\columnwidth]{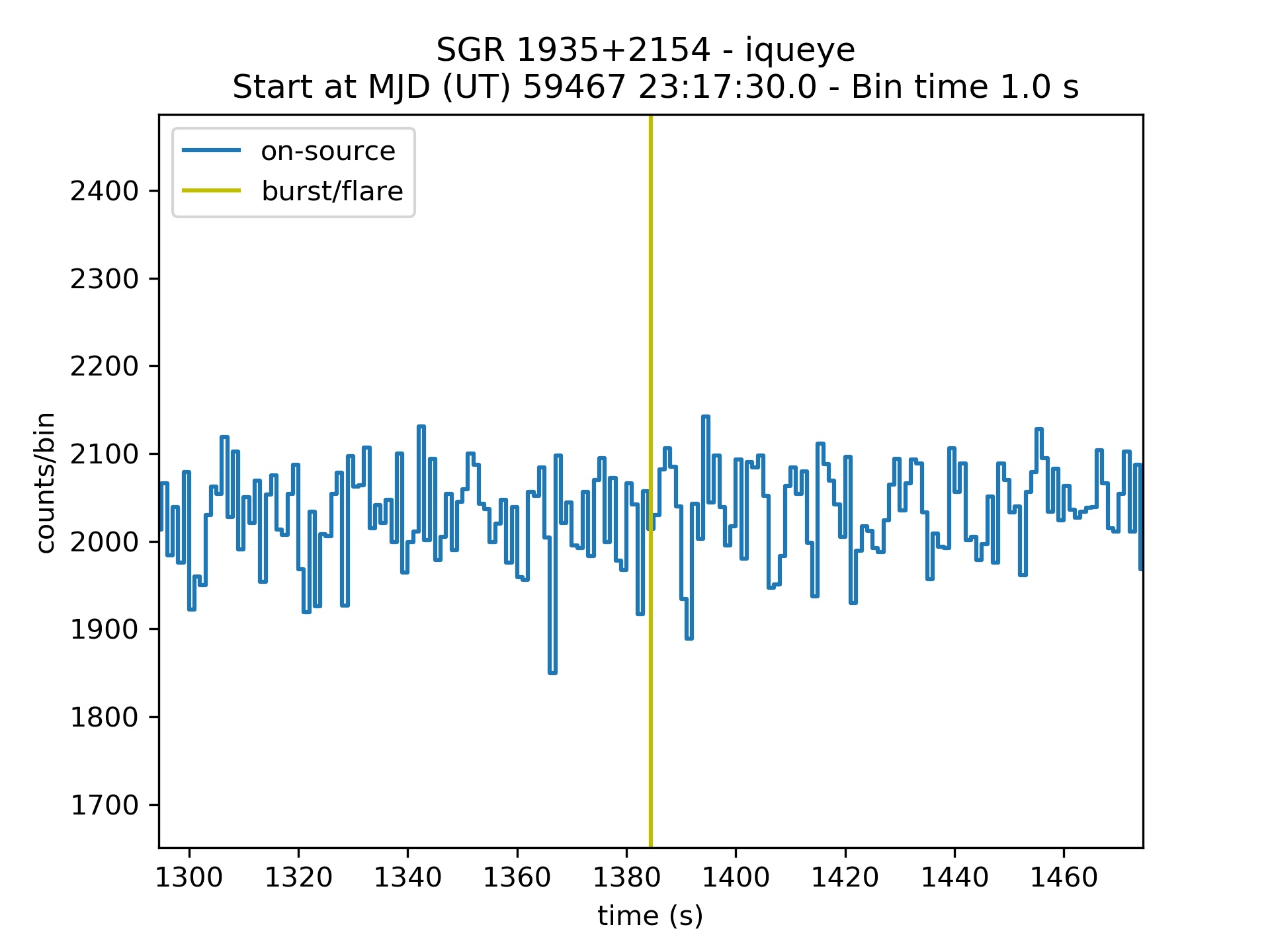}
        \includegraphics[angle=0, width=0.94\columnwidth]{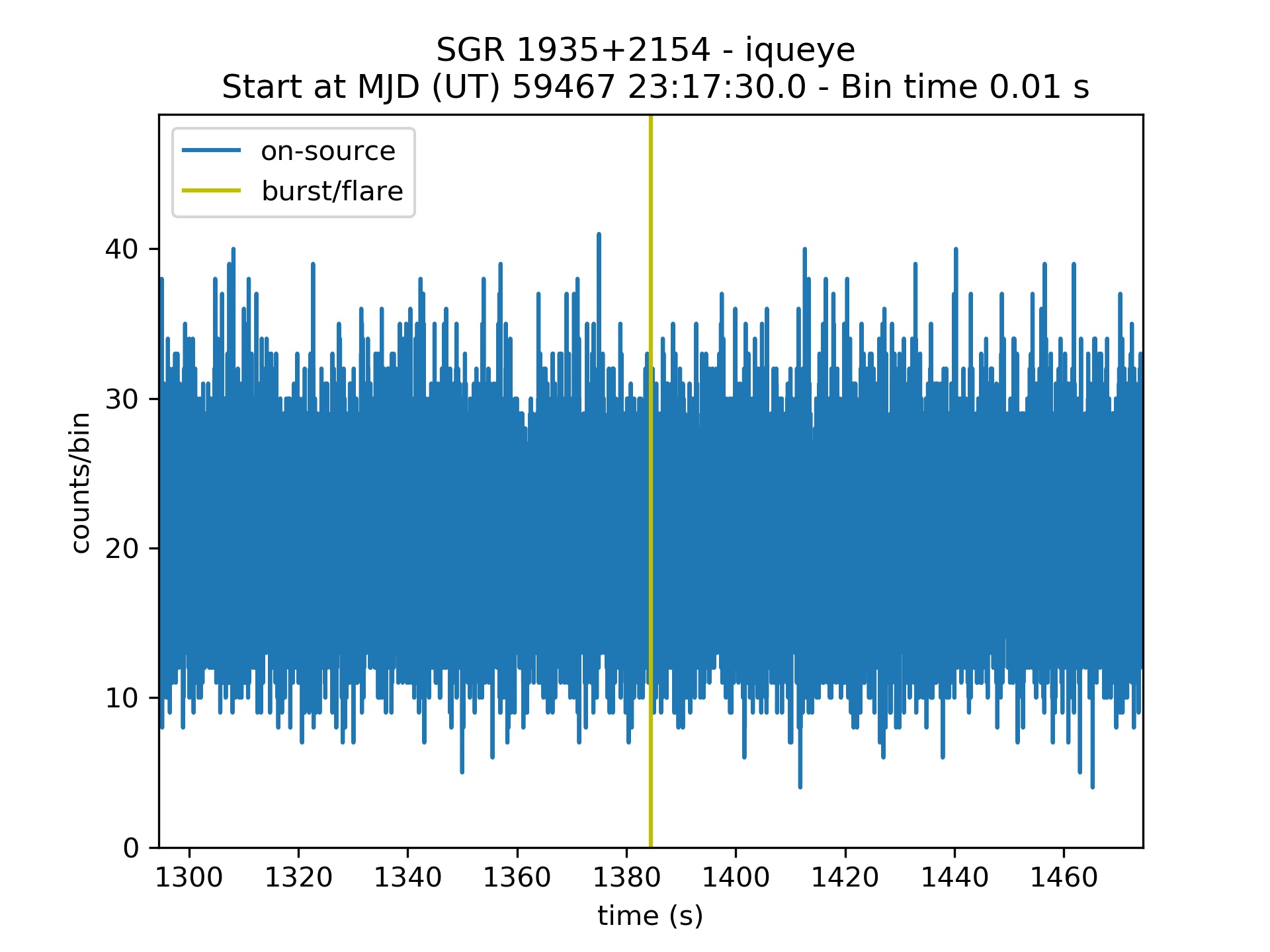}
	\includegraphics[angle=0, width=0.94\columnwidth]{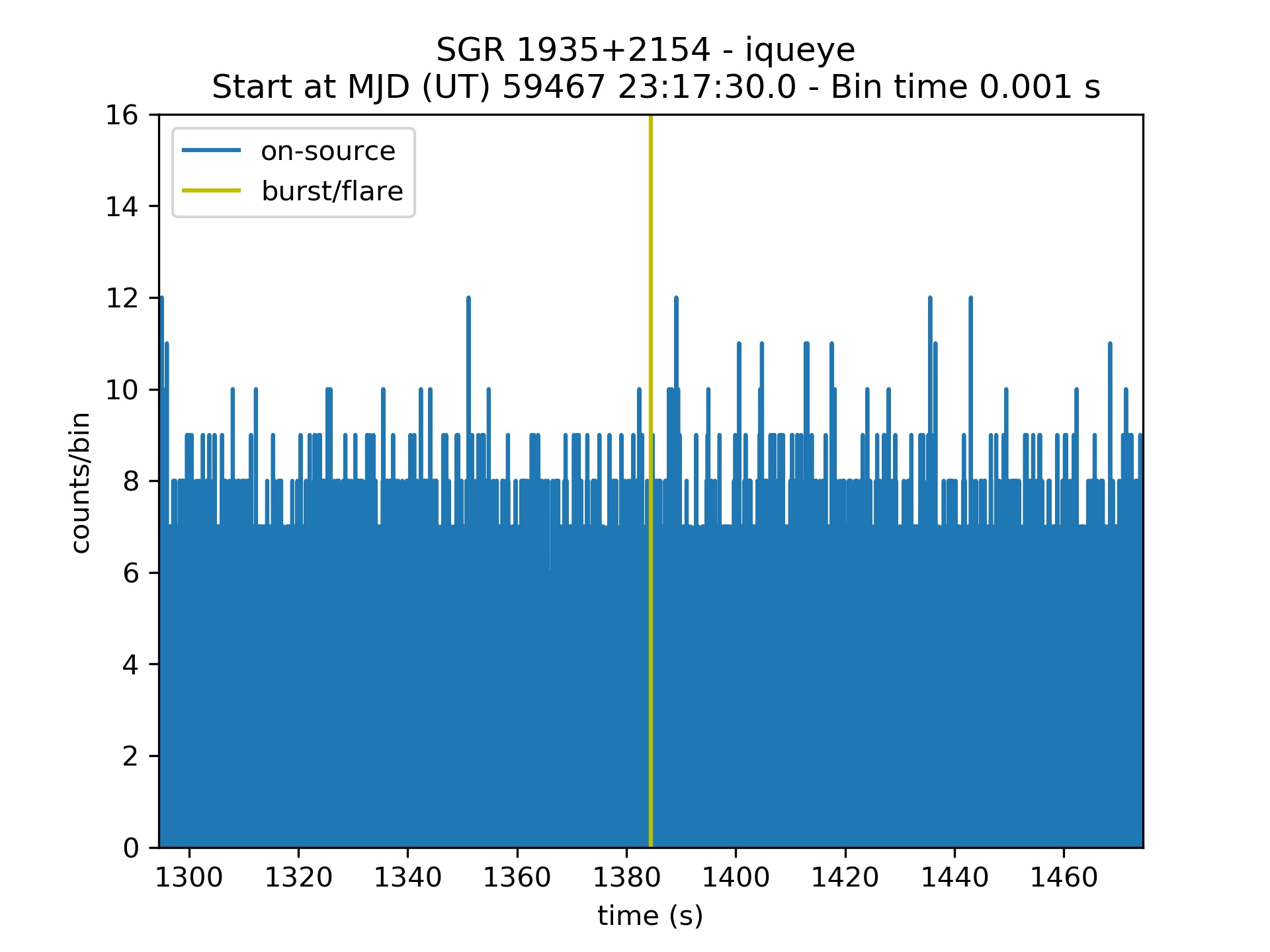}
    \caption{Zooming-in ($\pm$ 90 s) of the 1-s, 10-ms and 1-ms binned light curves of the second IFI+Iqueye observation of SGR J1935+2154 performed on Sep 10, 2021 (Table~\ref{tab:log}, Obs. ID 20210911-011727). The yellow line marks T$_0$, the time of arrival of the Fermi GBM burst. No significant peak is detected around T$_0$.}
    \label{fig:lciqz}
\end{figure}

We assume a Poisson distribution with the average rate of 2030 counts/s and fix a 3$\sigma$ detection threshold $n_t$ corresponding to a chance probability of $0.0027/N_{\rm trials}$ in any of the bins during an interval of $\pm$90 s around T$_0$. $N_{\rm trials}$ is the total number of bins in the interval and depends on the bin size. We obtain $n_t = 2220, 47, 14$ counts/bin for the 1-s, 10-ms and 1-ms binned light curves, respectively. No significant peak was detected above $n_t$ in any of the light curves (see Figure~\ref{fig:lciqz}).

The highest peak in the 1-s binned light curve (2140 counts/bin) translates into an upper limit to the signal of 230 counts/bin at the 3$\sigma$ confidence level, after subtracting a rate per bin that has a high Poissonian probability (99.73\%) to be exceeded by chance in a single bin (1910 counts/bin).
Using the $V$ band calibration of IFI+Iqueye \citep{2016SPIE.9907E..0NZ} and taking into account for the actual fiber-coupling efficiency at the time of the observation (59\%), a value of 230 counts/bin corresponds to an average optical brightness over 1 s of $V=15.9$ mag. Similarly, for the 10-ms and 1-ms binned light curve, the upper limits to the signal in the same interval are 32 counts/bin and 12 counts/bin (3$\sigma$ c.l.), respectively, corresponding to an average optical brightness of $V=13.1$ mag and $V=11.6$ mag.

We performed also a blind search for any significant increase in the count rate of the 1-s binned light curves of all the IFI+Iqueye observations. During each observation the count rate changes significantly because the sky diffuse background light increases with decreasing altitude of the target. For this reason, the blind search was performed after removing the increasing-background trend from the data. The upper limits to the signal in the 1-s binned detrended light curves of the three observations are: 290 counts/bin for observation 20210911-002926, 320 counts/bin for observation 20210911-011727, 290 counts/bin for observation 20210911-014851. The corresponding limits to the optical magnitudes are: $V=15.7$ mag, $V=15.6$ mag and $V=15.7$ mag, respectively.

The limits reported above refer to the apparent magnitude and do not take into account the extinction along the line of sight that, for SGR 1935+2154, is significant. We estimate the optical extinction in the direction of SGR J1935+2154 using the interstellar dust maps reported in \cite{2018JOSS....3..695G}\footnote{http://argonaut.skymaps.info/}, based on Pan-STARRS and 2MASS photometry and Gaia parallaxes of hundreds million stars \citep{2019ApJ...887...93G}. Using these maps and the distance estimate reported in \cite{2020ApJ...898L..29M} ($4.4_{-1.3}^{+2.8}$ kpc),
we find a reddening $E(g-r) \simeq 1.8$. The variation of $E(g-r)$ within most of the distance uncertainty range is small ($\lesssim 0.06$ mag). Only for a distance larger than 7 kpc, close to the upper boundary, the reddening increases significantly ($E(g-r) \simeq 2.4$).
From the transformation relations reported in \cite{2013MNRAS.430.2188Y}, we finally derive a $V$ band extinction of $A_V \simeq 5.8$ mag. This value is smaller but consistent within the uncertainties with the value $A_V \simeq 7.2 \pm 0.9$ mag \citep{2020ApJ...901L...7D} inferred from the neutral hydrogen column density along the line of sight obtained from XMM-Newton spectral fittings \citep{2016MNRAS.457.3448I}, although the latter is more uncertain and model-dependent.

Correcting for an extinction $A_V \simeq 5.8$ mag, the IFI+Iqueye upper limits to any possible optical burst from SGR J1935+2154 in an interval of $\pm$90 s around T$_0$ become $V=10.1$ mag for the 1-s binned light curve, $V=7.2$ mag for the 10-ms binned light curve, and $V=5.8$ mag for the 1-ms binned light curve.
The corresponding extinction corrected upper limits to the fluence (specific fluence) are $3.1 \times 10^{-10}$ erg cm$^{-2}$ (0.35 Jy s), $4.2 \times 10^{-11}$ erg cm$^{-2}$ (4.8 Jy $\cdot$ 10 ms), and $1.6 \times 10^{-11}$ erg cm$^{-2}$ (17.9 Jy ms).

\section{Discussion}
\label{sect:discussion}

\begin{figure*}
\begin{center}
        \includegraphics[angle=0, width=1.9\columnwidth]{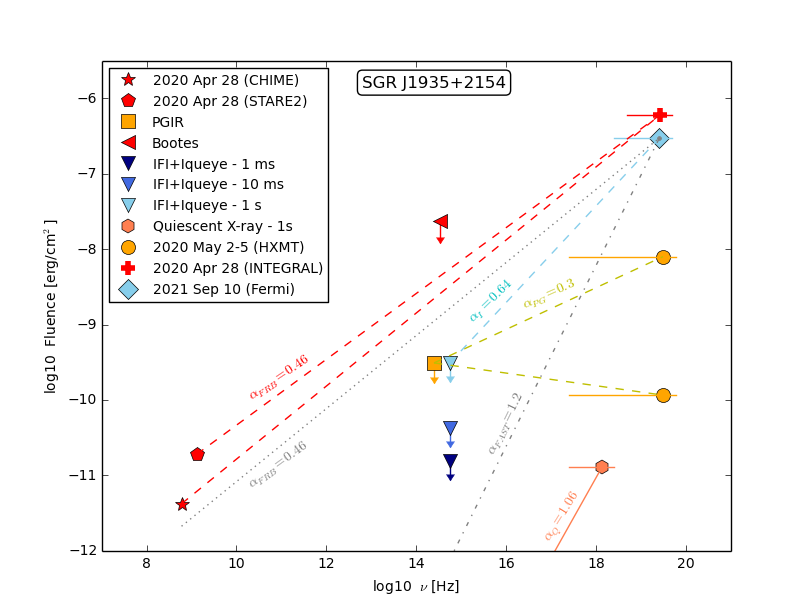}
        \caption{Upper limits to the optical fluence of SGR J1935+2154 during a simultaneous hard X-ray burst detected with Fermi/GBM. The IFI+Iqueye upper limits are extinction corrected and refer to the 1-s (upper cyan triangle), 10-ms (mid blue triangle) and 1-ms (lower dark blue triangle) binned light curves, respectively. The cyan dashed line shows the lower limit to the spectral index ($\alpha_I=0.64$) of the optical-gamma-ray fluence of SGR J1935+2154 assuming a power-law spectrum. Also shown are the radio and hard X-ray fluences of the 2020 April 28  burst (CHIME, STARE2 and INTEGRAL measurements), the bursts detected on 2020 May 2-5 (HXMT measurements), the upper limits (extinction corrected) derived from simultaneous observations in the optical and near infrared with BOOTES and Palomar-Gattini, respectively, and the corresponding fluence slopes or lower limits ($\alpha_{FRB}=0.46$ and $\alpha_{PG}=0.3$, red and yellow dashed lines). The gray lines represent hypotethical power-law dependences for the fluence of the 2021 Sep 10 event assuming a ``2020 Apr 28''-type burst ($\alpha_{FRB}=0.46$, dotted line) and a ``FAST''-type burst ($\alpha_{FAST}=1.2$, dot-dashed line; see text for details). Finally, the orange hexagon and the oblique orange line show the X-ray de-absorbed pre-outburst fluence (Swift measurement, integrated over 1 s) and the fluence slope ($\alpha_{Q}=1.06$) of the counterpart of SGR J1935+2154, calculated using the extinction corrected HST infrared fluence in a quiescent phase.}
    \label{fig:fluence}
\end{center}
\end{figure*}

The upper limits to the fluence of SGR J1935+2154 measured with IFI+Iqueye and the fluence of the simultaneous hard X-ray burst detected with the Fermi GBM are shown in Figure~\ref{fig:fluence} as a function of frequency. For comparison, we also report the measurements of the fluence of the radio and X-ray burst detected on 2020 April 28 \citep{2020Natur.587...54C,2020Natur.587...59B,2020ApJ...898L..29M}, the fluence of two other X-ray bursts detected few days later (2020 May 2-5) with HXMT \cite{2020Natur.587...63L}, and the X-ray de-absorbed fluence (Swift measurement, integrated over 1 s; \citealt{2016MNRAS.457.3448I}) of the quiescent (pre-outburst) counterpart of SGR J1935+2154.

We note that the radio to hard X-ray fluence ratio for the 2020 April 28 burst from SGR J1935+2154 is $\approx 10^{-5}$, while the limits for extragalactic FRBs imply a ratio $\gtrsim 10^{-(9 \div 10)}$ (\citealt{2021Univ....7...76N}; see also the limit of $> 10^{-9}$ for the closest extragalactic FRB in M81, \citealt{2021ApJ...921L...3M}).
Both on Apr 28 and May 2-5 there were simultaneous observations in the optical or near infrared bands at low time resolution ($\sim$1 s or larger): with the BOOTES telescope \citep{2020Natur.587...63L} the former and with the Palomar-Gattini telescope \citep{2020ApJ...901L...7D} the latter (see again Figure~\ref{fig:fluence}).

The results presented here appear to fit well within the framework of previous measurements and, at the same time, provide more stringent constraints on any possible optical impulsive counterpart to the SGR J1935+2154 bursts. Interestingly, the 2020 April 28 burst and that revealed with Fermi have comparable fluences. However, the 1-s IFI+Iqueye upper limit (extinction corrected) is two orders of magnitude deeper than the previous limit obtained with the BOOTES telescopes. We also note that the BOOTES limit was obtained with an integration time of 60 seconds. Our inferred limit to the optical to gamma-ray fluence ratio is $R_{\gamma, {\rm opt}} \lesssim 10^{-3}$. This value is consistent with the extinction-corrected fluence ratio observed in SGR J1935+2154 and other X-ray pulsars during long-term ($\sim$days-weeks) correlated X-ray/near-infrared outbursts ($R_{\rm X, NIR} \sim 10^{-4}$, 
e.g. \citealt{2018ApJ...854..161L}).


The mechanism producing magnetar bursts is still debated. They can be associated to the fast acceleration of magnetospheric particles after spontaneous magnetic field reconnection
(see \citealt{2003MNRAS.346..540L,2005ApJ...629..985W}; see also \citealt{2016MNRAS.456.3282E}), 
or to emission from a magnetically confined pair fireball, driven by the conversion into a hot, optically thick  electron-positron plasma of large-amplitude magnetospheric oscillations \citep{1995MNRAS.275..255T,2001ApJ...561..980T}. The spectral properties of the burst radiation have been investigated in some detail for the latter scenario, although a number of open issues still remain \citep{2002MNRAS.332..199L,2015ApJ...815...45Y,2016MNRAS.461..877V,2017MNRAS.469.3610T}. Emission from the fireball itself is thermal, likely described by the superposition of two blackbodies, at an effective temperature of $\approx 10$ keV, in agreement with observations (see e.g. the analysis of the ``burst forest'' of SGR J1900+14 by \citealt{2008ApJ...685.1114I}). Accordingly, no low-energy emission should be expected in a burst. However, energetic photons from the hot pair plasma heat the neutron star surface, ablating material which is then carried away in the form of a baryon-rich outflow, as first pointed out by \cite{2001ApJ...561..980T} (see also \citealt{2016MNRAS.461..877V}), giving rise to a much more complicated picture. 

It has been recently suggested that the ultrarelativistic blast wave produced by the interaction of the plasmoid ejected in a magnetar (giant) flare with the magnetar wind in the tail of a previous flare can result in an optical flash via (incoherent) synchrotron emission \citep{2020ApJ...896..142B,2020ApJ...893...99C}.
The energetics of the burst considered here is much below that of a giant flare, so it is unclear whether these considerations still hold. However, by applying to SGR J1935+2154 the scaling between the optical and the total energy released in the burst proposed by \cite{2020ApJ...896..142B}, the expected optical fluence is $\approx 10^{-2}$--$10^{-4}$ the high energy one, depending on the magnetization parameter of the wind. Our present upper limit,  $R_{\gamma,  {\mathrm{opt}}}\sim 10^{-3}$, would be still compatible with such a picture, possibly pointing to a somehow large magnetization.

The IFI+Iqueye measurement can also place a more stringent constraint to the spectral index of the optical to gamma-ray fluence of SGR J1935+2154. Assuming hereafter a single power-law dependence for the fluence $\tilde{F}_\nu \propto \nu^\alpha$, the index has to be larger than $\alpha_I = 0.64$ to be consistent with the optical non detection (see Figure~\ref{fig:fluence}).
Interestingly, the optical limit lies above the extrapolation of the GBM spectral fit (for which $\alpha = 2-\Gamma = 0.78 \div 1.5$, where $\Gamma = 0.5 \div 1.22$; see Sect.~\ref{sect:fermi}).
While the Palomar-Gattini near infrared fluence limit is similar to that obtained with IFI+Iqueye, the simultaneous X-ray burst detected with HXMT was weaker than the Fermi one and the inferred limit on the slope of the fluence is less constraining ($\alpha_{PG} = 0.3$, \citealt{2020ApJ...901L...7D}). Similarly, the BOOTES upper limit is not sufficiently deep to even rule out a single power-law with the spectral index calculated from the simultaneous radio-gamma-ray detections of FRB 200428 ($\alpha_{FRB} = 0.46$).
For comparison, we also report the fluence slope ($\alpha_{Q}=1.06$) of the counterpart of SGR J1935+2154, calculated using the de-absorbed pre-outburst Swift X-ray fluence ($1.28 \times 10^{-11}$ erg cm$^{-2}$, \citealt{2016MNRAS.457.3448I}) and the extinction corrected quiescent Hubble Space Telescope (HST) infrared fluence ($1.19 \times 10^{-15}$ erg cm$^{-2}$, \citealt{2018ApJ...854..161L}), both integrated over 1 s. For this purpose, we neglected the long timescale variability of the infrared source and considered only the dimmest reported measurement \citep{2018ApJ...854..161L,2021arXiv211207023L}. The fluence slope of the counterpart is somewhat steeper than the one that IFI+Iqueye can probe in the case optical outbursts of comparable magnitude accompany the X-ray bursts.

Evidence for a steep spectral index for some bursts was also reported by \cite{2020Natur.587...63L}. They found that 29 bursts from SGR J1935+2154 detected with Fermi GBM were not simultaneously detected in the radio band with FAST. The inferred limit to the radio fluence for these bursts imply a rather steep spectral index for the radio to gamma-ray fluence ($\alpha_{FAST} = 1.2$, \citealt{2020Natur.587...63L}) for the majority of the bursts from SGR J1935+2154.
This fact led \cite{2020ApJ...901L...7D} to conclude that the Palomar-Gattini near infrared measurements would not be sufficiently deep to detect possible counterparts of the majority of the bursts.

If we assume that there are two types of bursts in SGR J1935+2154, the rarer ``2020 Apr 28''-type bursts and the more common ``FAST''-type bursts, and that they are characterized by a single power-law dependence for the fluence, there are two important consequences for the reported measurements. First, the IFI+Iqueye non-detection of the Fermi Sep 10, 2021 burst is consistent with the ``FAST''-type event, as the FAST limit on the fluence slope ($\alpha_{FAST} = 1.2$) is steeper than that from IFI+Iqueye ($\alpha_I = 0.64$) and hence the extrapolation of the FAST power-law dependence is expected to fall below the 1-s optical upper limit (dot-dashed line in Figure~\ref{fig:fluence}). Indeed, if the majority of the bursts from SGR J1935+2154 are of the ``FAST''-type, they will not be detectable with IFI+Iqueye. Conversely, in the same assumptions, a smaller number of ``2020 Apr 28''-type bursts, characterized by a much flatter radio-through-hard-X-ray slope ($\alpha_{FRB} = 0.46$), are in principle detectable in the optical band with a simultaneous IFI+Iqueye observation, as the extrapolation of the power-law dependence for the 2020 Apr 28 fluence is above the 1-s IFI+Iqueye upper limit (dotted line in Figure~\ref{fig:fluence}). Therefore, future observations with fast photometers like IFI+Iqueye, simultaneous with ``2020 Apr 28''-type radio+X-ray bursts, have the potential to yield an optical detection. Even if this type of bursts is rare, these findings represent a strong motivation for undertaking dedicated fast optical timing campaigns during future outburst episodes of SGR J1935+2154.

Finally, we note that, as shown in Figure~\ref{fig:fluence}, the IFI+Iqueye upper limit to the fluence for 10-ms and 1-ms optical bursts is a factor 7 and 20, respectively, deeper than that of the 1-s binned light curve. While these are not the typical durations of the bursts from SGR J1935+2154, in the future it will be interesting to investigate what limits can be derived by comparing the fluence of 1-10 ms substructures in the radio, X-ray and gamma-ray bursts with simultaneous fast optical timing observations on these (or even shorter) timescales.

\acknowledgments

We thank the referee for useful and constructive comments. L.Z. acknowledges financial support from the Italian Space Agency (ASI) and National Institute for Astrophysics (INAF) under agreements ASI-INAF I/037/12/0 and ASI-INAF n.2017-14-H.0 and from INAF ``Sostegno alla ricerca scientifica main streams dell'INAF'' Presidential Decree 43/2018. S.M. and R.T. are partially supported by the Italian MIUR through grant ``UNIAM'' (PRIN 2017LJ39LM). E.P.  and  A.P. acknowledge support from PRIN-MIUR 2017 (grant 20179ZF5KS).
Based on observations collected at the Galileo telescope (Asiago, Italy) of the University of Padova. This research has made use of data and software provided by the High Energy Astrophysics Science Archive Research Center (HEASARC), which is a service of the Astrophysics Science Division at NASA/GSFC.


%

\vspace{5mm}
\facilities{Galileo Telescope Asiago(IFI+Iqueye), Fermi(GBM)}


\software{MATPLOTLIB \citep{2007CSE.....9...90H},  
          NUMPY \citep{2011CSE....13b..22V}
          ASTROPY \citep{2018AJ....156..123A}
          ASTROQUERY \citep{2019AJ....157...98G}
          }

\end{document}